# Upstream neutral modes in the fractional quantum Hall effect regime: heat waves or coherent dipoles

Yaron Gross, Merav Dolev, Moty Heiblum, Vladimir Umansky, and Diana Mahalu

Braun Center for Submicron Research, Dept. of Condensed Matter Physics,

*Weizmann Institute of Science, Rehovot* 76100, *Israel*

**ABSRACT**

**Counter propagating (upstream) chiral neutral edge modes, which were predicted to be present in hole-conjugate states, were observed recently in a variety of fractional quantum Hall states ($\nu=2/3$, $\nu=3/5$, $\nu=8/3$ & $\nu=5/2$), by measuring charge noise that resulted after partitioning the neutral mode by a constriction (denoted, as N→C). Particularly noticeable was the observation of such modes in the $\nu=5/2$ fractional state - as it sheds light on the non-abelian nature of the state's wavefunction. Yet, the nature of these unique, upstream, chargeless modes and the microscopic process in which they generate shot noise, are not understood. Here, we study the ubiquitous $\nu=2/3$ state and report of two main observations: First, the nature of the neutral modes was tested by 'colliding' two modes, emanating from two opposing sources, in a narrow constriction. The resultant charge noise was consistent with local heating of the partitioned quasiparticles. Second, partitioning of a downstream charge mode by a constriction gave birth to a dual process, namely, the appearance of an upstream neutral mode (C→N). In other words, splitting 'hole conjugated' type quasiparticles will lead to an energy loss and decoherence, with energy carried away by neutral modes.**

Quantum Hall states are characterized by the chiral flow along the edges of a two dimensional electron gas (2DEG) of one-dimensional-like current modes [1,2]. For some fractional states (such as hole-conjugate states, *e.g.*, $v=2/3, v=3/5$, etc.), MacDonald *et al*. [3,4] speculated the existence of counter propagating ('upstream') charge modes due to 'edge reconstruction'; however, such modes were never observed [5]. This motivated Kane *et al*. [6,7] to propose formation of upstream neutral modes, which carry energy without net charge, due to Coulomb interaction and particle exchange between the proposed counter propagating charge modes. The recent observation of neutral modes by Bid *et al.* [8], via shot noise measurements, had been accomplished by impinging upstream neutral modes on a partly pinched quantum point contact (QPC) constriction, opened a new field of study. Here, we study in some detail the birth of the modes in the ubiquitous $v=2/3$ state and the mechanism by which they generate shot noise in a constriction.

The $v=2/3$ mode was verified recently to be made of a downstream charge mode with conductance $2e^2/3h$ and an upstream neutral mode carrying only energy [6,7,8]. When tunneling between two counter propagating 2/3 modes takes place (such as in a narrow constriction), the measured tunneling quasiparticles were found to be $e/3$ or $2e/3$ [9], approaching $e$ for a nearly pinched constriction. Moreover, the simultaneous presence of an excited neutral mode in the constriction was found to add charge fluctuations and, at the same time, modify substantially the downstream flowing partitioned quasiparticle charge [8,10]. However, we have yet to understand many of these modes' characteristics, and some of them are studied here. What is the role played by the ohmic contact in emitting and absorbing these modes? What is the microscopic mechanism that leads to current noise in a narrow constriction? [8,10] Since charged quasiparticles and neutral ones coexist, it is only natural to ask whether a dual behavior, namely, partitioning charged quasiparticles will excite upstream neutral modes. Moreover, and most importantly, can the neutral modes be described as an ordered stream of neutral quasiparticles with clear quantum numbers and statistics (say dipoles, meaning a spinor), or they are propagating incoherent heat waves.

Measurements were performed on samples fabricated on a GaAs-AlGaAs heterostructure, embedding, some 130*nm* below the surface, a two dimensional electron gas (2DEG) with areal carrier density ~$9\times10^{10}cm^{-2}$ and low temperature (<1*K*) mobility ~$6\times10^{6}cm^{2}/V\text{-}s$ - both measured in the dark. A schematic representation of the fabricated structure is shown in Figs. 1a, 1b and Figs. 4a & 4b. It is composed of a mesa 300*μm* long and 50*μm* wide, with a 10*μm* narrow part covered by a split metallic gate, 100*nm* wide and opening ~450*nm*. Alloying AuGeNi in the usual manner formed ohmic contacts. Due to the relatively short decay length of the neutral modes (see Ref. 8), the distance between the 'injecting' ohmic contacts and the detector (e.g., a quantum point contact (QPC)) was kept relatively short (8*μm*). The signal at the voltage probe was first filtered by LC circuit, with a resonance frequency ~850*kHz*, amplified by a cooled (to 4.2*K*) home made pre-amplifier and a room temperature amplifier (NF-220F5), to be finally measured by a spectrum analyzer. Grounded contacts were connected directly to the 'cold finger' of a dilution refrigerator at temperature ~10*mK*, assuring electron temperature in this range.

We first address the nature of the neutral mode and the process by which it generates charge noise at the QPC (we denote this process N→C) [8,10]. A modified experiment was designed, where rather than having a single upstream neutral mode impinging at a constriction with a resultant downstream charge noise $S_{single}$, two neutral modes were injected simultaneously, from opposite sides of the constriction, with a resultant charge noise $S_{collision}$ (see schematic in Fig. 1b). The experiment proceeded as follows: First, current was injected from a single contact, and $S_{single}$ for $I_{N1}$ (or $S_{single}$ for $I_{N2}$) was measured at N2. Both configurations led to very similar results. Second, currents were injected from $N_1$ and $N_2$ simultaneously, generating $S_{collision}$ at N2. The measurements were repeated for different transmission probabilities set by the QPC. Note, that since N2 has two roles, injecting and collecting, a large resistor (1*Gohm*) was inserted directly on the N2 contact; avoiding shorting the noise by the stray capacitance of the wire feeding N2.

Our findings, shown in Figs. 1c & 1d, indicate that for all the values of the transmission probability $S_{single}<S_{collision}<2S_{single}$; moreover, simple scaling of the graphs allow expressing $S_{collision}=\alpha S_{single}$, with $\alpha=1.6\pm0.1$ (Figs. 1e & 1f). Both spectra, $S_{single}$ and $S_{collision}$, exhibited the same dependence on the transmission probability, $t(1-t)$, as shown in Fig. 2a. Note that these results stand in contrast to experiments where two unpartitioned (noiseless) charge modes collide from both sides at a constriction with a null resultant noise ($S_{collision}=0$). This is a trivial outcome of two full Fermi seas impinging from both sides resulting also with two outgoing full Fermi seas.

The most straightforward explanation for $S_{single}$ ($S_{collision}$) is local heating of the counter propagating 'cold' charge mode(s) by the neutral mode(s), which results in an elevated temperature in the constriction [11,12], thus generating excess charge noise above the Johnson-Nyquist noise in the absence of neutral mode(s) [13,14]. Since, to the best of our knowledge, the consequence of a non-uniform temperature along the sample is not available, we derive a simplified description of such unorthodox case. The derivation follows Landauer's guidelines [15,16] applied to the configuration in Fig. 2b. The Fermi-Dirac distributions in contacts C1, C2 and the ground G1, are: $f_1(\mu_1,T_1)$, $f_2(\mu_2,T_2)$, $f_g(\mu_g,T_g)$, respectively, with $\mu$ the electrochemical potential and $T$ the temperature. The current distribution injected from a contact is proportional to $fg_Q$, with $g_Q$ the Hall conductance, whilst its variance is proportional to $f(1-f)g_Q$. Since the voltage fluctuations in N2 are with respect to ground G1, it is the sum of the variances in these contacts. Since the current distribution arriving at N2 is proportional to $F_{N2}=f_1t+f_2(1-t)$, and that leaving G1 is proportional to $f_g$, the measured excess noise at N2 is given by:

$$S_{N2} = 2g_Q \int_0^\infty \{F_{N2}(1-F_{N2}) + f_g(1-f_g)\} dE \quad . \tag{1}$$

This expression can be divided into two contributions: $S_{contacts}$ – the spectrum density emanating from reservoirs and transmitted (or reflected) by the constriction as $t^2$ (or $(1-t)^2$), and $S_{QPC}$ – the spectrum density that results in the QPC constriction due to partitioning.

$$S_{contacts} = 2g_Q \int_0^\infty [\ f_1(1-f_1)t^2 + f_2(1-f_2)(1-t)^2 + f_g(1-f_g)\ ] dE$$
$$S_{QPC} = 2g_Q \int_0^\infty \{\ t(1-t)[f_1(1-f_2) + f_2(1-f_1)]\ \} dE \tag{2}$$

In thermal equilibrium, with no voltage applied, $T_1=T_2=T_g$ and $\mu_1 = \mu_2$:

$$\begin{aligned} S_{contacts} &= 4k_B T g_Q (t^2 - t + 1) \\ S_{QPC} &= 4k_B T g_Q\, t(1-t) \end{aligned}, \tag{3}$$

With the total $S=S_{contacts}+S_{QPC}$ measured in N2 is, as expected, $S_{thermal} = 4k_b T g_Q$ (independent of $t$). The excess noise, induced by increasing the electrochemical potential of the injected edge is:

$$S_{excess} = 2\Delta\mu g_Q t(1-t)\left[\coth\left(\frac{\Delta\mu}{2k_B T}\right) + \frac{2k_B T}{\Delta\mu}\right] \quad . \tag{4}$$

When the temperature is not the same in all contacts, their contribution to the noise is:

$$S_{contacts} = 2g_Q k_B \left(T_1 t^2 + T_2(1-t)^2 + T_g\right), \tag{5}$$

however, that of the single constriction ($S_{QPC}$) can be solved only numerically.

We apply now this model to local heating in the constriction. We assume that the neutral mode increases the temperature of the current carrying channel in one side of the constriction from $T_g$ to $T_n$, while all other contacts remain at $T_g$, since the neutral mode decays before reaching the contacts). Such energy exchange may be the reason for the decay of the neutral mode as it propagates upstream. The numerical solution for $T_n \gg T_g$ results in a modified Johnson-Nyquist noise,

$$S_{QPC-single} \simeq 2.75 k_b T_n g_Q t(1-t) \ . \tag{6}$$

Evidently, when the two sides of the constriction are heated to $T_n$ in the 'collision' type experiment, the Johnson-Nyquist noise is,

$$S_{QPC-collision} = 4 k_B T_n g_Q t(1-t) \ , \tag{7}$$

agreeing with the main experimental observations for N→C as follows:

1. The temperature $T_n$ grows linearly with the injected current, in agreement with the Wiedemann-Franz law (see Ref. 11), hence also the observed noise ($T_n \gg T_g$), reaching $T_n \sim 250$mK for $I_n=2$nA;

2. Observed $S_{QPC-collision} = \alpha \, S_{QPC-single}$, with $\alpha = 1.6 \pm 0.1$, while estimated $\alpha = 1.45$;

3. Measured and projected noise follow $t(1-t)$ dependence;

While the agreement with measurements is encouraging, the temperature estimate it provides ($T_n \sim 250$mK) stands in contrast with measurements where a charge mode $I_{C1}$ was injected from C1, and a neutral mode emanated from N1 (driven by current $I_{n1}$; with the two modes arriving at two opposite edges of the constriction. While the 'charge side' of the constriction is at $T_g$ the 'neutral side' is at $T_n$. For a constant $I_{n1}$, our model predicts the two contributions to the excess noise are: an increased thermal noise for all $I_{C1}$ with $T \to T_n/1.45$, and shot noise due to the partitioned $I_{C1}$ obeying Eq. 4 with $T \to T_n/1.45$. The measured excess noise at $I_{C1}=0$ and $I_{n1}=2$nA suggests $T_n \sim 250$mK (as above) while the dependence of the excess noise on $I_{C1}$ (the 'rounding' near $I_{C1}=0$, [17]) leads to $T_n \sim 50$mK accompanied by an unpredicted reduction of the tunnelling charge in the constriction (see Fig. 2c) [see also Ref. 8]. While the reason for this discrepancy is not clear, recalling that a similar reduction from $2e/3$ to $e/3$ takes place with increasing the temperature to ~100mK [9] may support the higher temperature estimate.

The interplay between charge and neutral modes raises the obvious question whether partitioning of a charge mode results also with energy transfer to an excited upstream neutral mode (C→N). In principle, in order to perform this experiment one would need two constrictions; the first to test the C→N process and the second to detect the presence of the neutral mode via N→C process. However, and unexpectedly, and contrary to the previous experiments [8], we found that all our ohmic contacts generated noise (with no net current) when absorbed an upstream neutral mode at ν=2/3, ν=3/5, ν=5/2, ν=5/3, and ν=8/3 (in different samples) as shown in Fig 3. This effect may be related to the unique location of the heat dissipation we impose at the 'front side' of the ohmic contact. Usually, a 'hot spot' is generated at the "back side" of a charged contact [18], when a difference in electrochemical potentials leads to dissipation with temperature increase. Here, an upstream neutral mode arrives at the "front side" of an unbiased contact, and thus heats up the injected downstream charge mode. The distribution $f(\mu=0, T_g)$ changes to $f(\mu=0, T_n)$, resulting in excess noise $2k_B g_Q (T_n - T_g)$. For the current configuration, where the injecting and detecting contacts are 25μm apart, the measured noise suggests $T_n \sim 40 mK$ - not an unreasonable temperature. It is presently unclear why this effect was not observed in the previous samples [8]. As we study this effect, we can only speculate now that it crucially depends on details in the interface between the contact and the 2DEG.

An injected downstream current $I_{C1}$ was partitioned by a constriction with a resultant upstream neutral mode, which was detected at M1 (C→N process, see Fig 4b). Like in the N→C process, the noise increased with source current with a dependence $t(1-t)$ (Figs. 4d & 4f). Being the ubiquitous dependence of shot noise on $t$ due to stochastic partitioning, it suggests that the neutral mode is an outcome of the stochastic tunneling process of charged quasiparticles in the constriction. The observation of the C→N process opens up another way to identify the ground state wavefunction of the $\nu=5/2$ state – the current best candidate for a non-abelian state [21]. It might be important to note that such excitation of neutral modes may render the partitioning of charged quasiparticles inelastic, leading to incoherent scattering by the constriction, limiting its use as an elastic 'beam splitter' in interference experiment [19, 20].

Here, we explore in some depth some of the very basic properties of the upstream neutral modes in the 2/3 fractional state. We focused mainly on the mechanism of their detection and the interplay between the charge and neutral modes in a constriction. Our results suggest that neutral modes are likely to excite charge modes, in the narrow constriction, by heating them up. We observed a dual process, in which, partitioned charged quasiparticles transfer energy to upstream neutral modes. This observation brings to light an important question: can partitioned charged quasiparticles, in 'hole conjugate' fractional state, split, and thus partition, elastically, as needed in interference experiments.

**Acknowledgments**

We thank Bernd Rosenow, Ariel Amir, Ady Stern, Yuval Gefen, Charles Kane, Eytan Grosfeld and Steve Kivelson, for helpful discussions. We acknowledge the partial support of the European Research Council under the European Community's Seventh Framework Program (FP7/2007-2013) / ERC Grant agreement # 227716, the Israeli Science Foundation (ISF), the Minerva foundation, the German Israeli Foundation (GIF), the German Israeli Project Cooperation (DIP), and the US-Israel Bi-National Science Foundation (BSF).

# References

[1]  S. Das Sarma, A. Pinczuk, Perspectives in Quantum Hall Effect. Wiley (1997).

[2]  X.G. Wen, Phys. Rev. B **41**, 12838 (1990).

[3]  A. H. MacDonald, Phys. Rev. Lett. **64**, 220 (1990).

[4]  M. D. Johnson, A. H. MacDonald, Phys. Rev. Lett. **67**, 2060 (1991).

[5]  R. C. Ashoori, H. L. Stormer, *et al*., Phys. Rev. B **45**, 3894 (1992).

[6]  C. L. Kane, M. P. A. Fisher, J. Polchinski, Phys. Rev. Lett. **72**, 4129 (1994).

[7]  C. L. Kane, M. P. A. Fisher, Phys. Rev. B **51**, 13449 (1995).

[8]  A. Bid, *et al*., Nature **466**, 585 (2010).

[9]  A. Bid, *et al.,* Phys. Rev. Lett. **103**, 236802 (2009).

[10]  M. Dolev, Y. Gross, *at el*., Phys. Rev. Lett. **107**, 036805 (2011).

[11]  S. Takei, B. Rosenow, arXiv:1012.0315.

[12]  E. Grosfeld, S. Das, Phys. Rev. Lett **102**, 106403 (2009).

[13]  J. Johnson, Phys. Rev. **32**, 97 (1928).

[14]  H. Nyquist, Phys. Rev. **32**, 110 (1928).

[15]  R. Landauer, Physica (Amsterdam) **38D**, 226 (1989).

[16]  R. Landauer, T. Martin, Physica B **175**, 167 (1991).

[17]  M. Reznikov, *et al.*, Superlattices and Microstructures **23**, 901 (1998).

[18]  U. Klass, *et al*., Z. Phys. B. **82**, 351 (1990)

[19]  A. Stern, I. Halperin, Phys. Rev. Lett. **94**, 016802 (2006).

[20]  P. Bonderson, *et al.,* Phys. Rev. Lett. **97**, 016401 (2006).

[21]  D. E. Feldman, F. Li, Phys. Rev. B **78**, 161304 (2008)

# Figure Captions

**Figure 1**

**'Collision' measurements - setup and results.** Schematics of the measurements setup is presented at **a)** & **b)**. Ohmic contacts in green, 2DEG in gray, and QPC constriction in black. Chiral neutral (charge) mode in red (blue); flowing in opposite directions. Distance from N1 and N2 to the center of QPC opening 8μm. Distance from C1 and C2 to center of QPC opening 150μm. **a)** N→C setup. Neutral mode injected (upstream) from source N1 toward the QPC, where it generates excess noise $S_{single}$, which flows (downstream) towards voltage probe N2. **b)** 'Collision' setup. Neutral mode injected (upstream) simultaneously from contact N1 & N2, flowing toward the QPC, where they generate excess noise $S_{collision}$, which is measured in voltage probe N2. **c) & d)** Excess noise generated when a single (red, $S_{single}$) or two (black, $S_{collision}$) neutral modes reach the QPC, for two different transmission probabilities $t$. $S_{single}$ does not depend on the injecting contact (N1 or N2). **e) & f)** Scaling $S_{collision}$; $S_{collision}/S_{single}=1.6$.

**Figure 2**

**'Collision' experiment – results and model.** **a)** Dependence of excess noise at injection current $2nA$ on transmission probability $t$ of the QPC - one (red triangles) or two (black squares) upstream neutral mode(s) reach the QPC. The noise follows $t(1-t)$ trend (represented by a dashed line), with $S_{collision}/S_{single}=1.6$ for all $t$. Empty/full shapes represent measurements performed on different samples. **b)** Schematics of the model calculation in the text. Charge mode propagates from C1 or C2 toward the voltage probe and partitioned by the QPC. The measured noise signal is made of the emitted noise from the contacts and the noise generated at the QPC. **c)** Noise generated by partitioning the charge mode. Without an excited neutral mode (blue), the noise agrees with tunneling of 2$e$/3 quasiparticles at low temperature. When a constant neutral mode is introduced to the opposite edge (red, $I_n$=2nA), it elevates its temperature to $T_n$, and a reduction in the tunneling quasiparticles charge (to $e$/3) is observed. A temperature increase to $T_n$~50mK is estimated from the dependence of the noise near $I_c$=0 as. In contrast, the increase in background noise leads to $T_n$~250mK.

**Figure 3**

**Detection of neutral mode by an ohmic contact.  a)** Schematics of the measurement setup. Injecting current from contact N3, gives birth to a neutral mode flowing upstream, reaching the voltage probe N2 (25μm away), where it generates excess noise. **b)** Noise at voltage probe N2 as function of injected current at N3 for ν=2/3 (black). No noise is observed at ν=1 (red). **c)** Similar results at the first excited Landau level, measured on a different sample. Noise signal is observed at ν=5/2 but not at ν=7/3.

**Figure 4**

**Duality of N→C and N→C.  a) & b)** Schematics of the measurement setup for N→C and C→N, respectively. In the C→N process, current (blue) is injected from source C1, flowing downstream and reaching the QPC, where it is partitioned. Neutral mode (red), generated at the QPC, flows upstream and reaches the voltage probe M1 (8μm apart). The distance from the center of the QPC to C1 is 150μm. **c) & d)** Noise measured at the voltage probe as function of the injected current for different transmission probabilities $t$ of the QPC. Noise in N→C process is presented in **c)**, and noise in C→N process is presented in **d)**. Data points are connected by a guide to the eye. **e) & f)** Noise measured at the voltage probe for different transmission probabilities *t* for 1*nA* injection current. In the two plots the data follows *t*(1-*t*) trend – given by a dashed line. Empty/full circles represent data point from different samples.

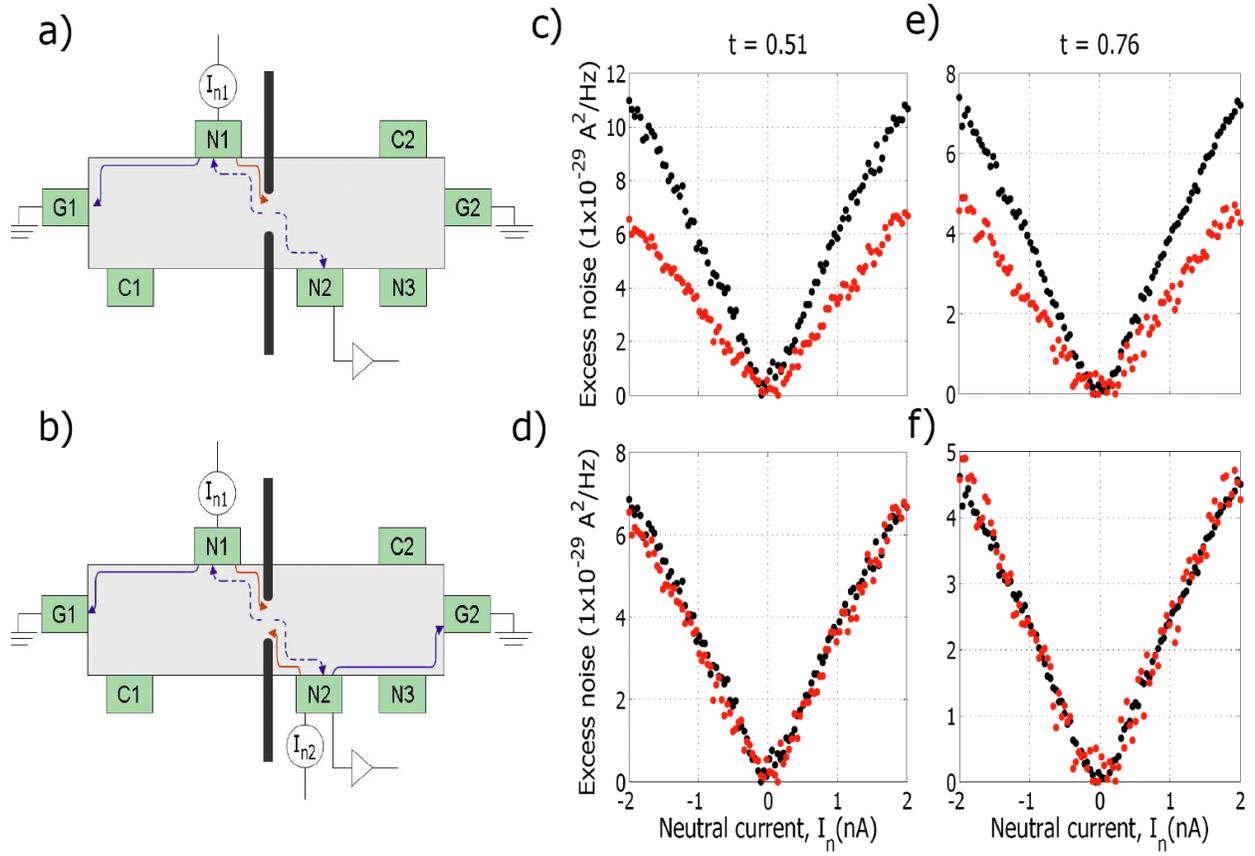

**Figure 1**

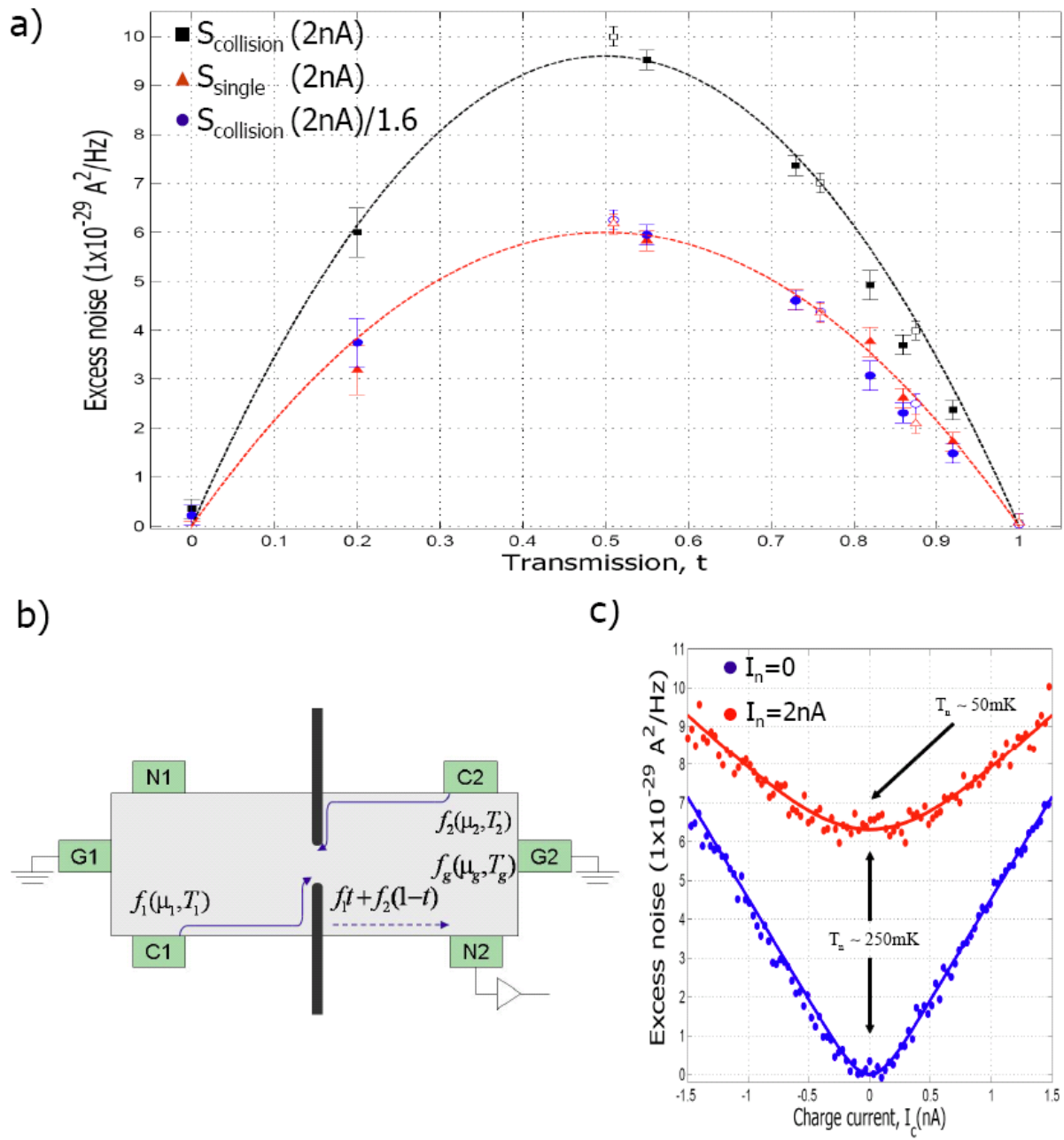

**Figure 2**

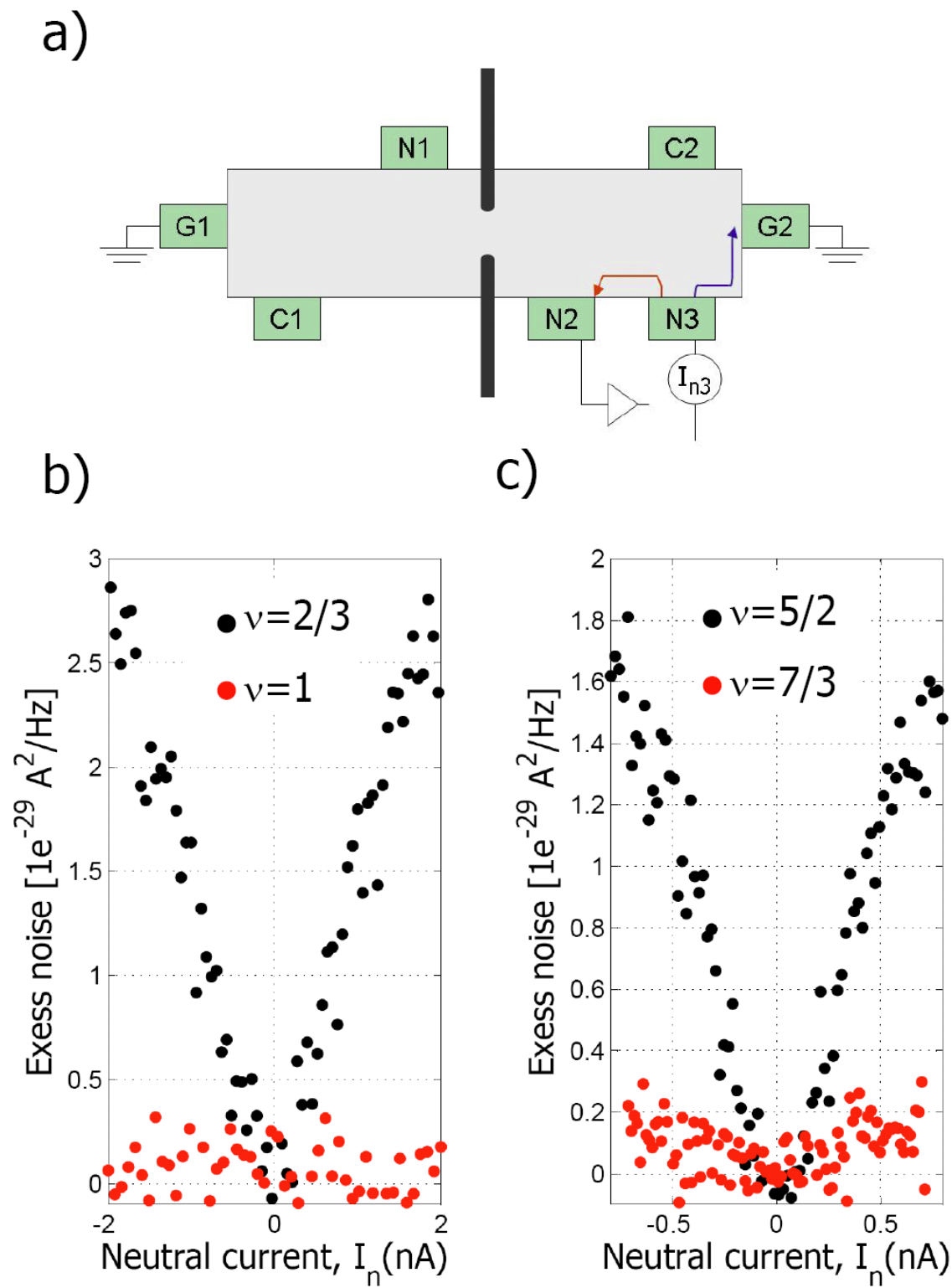

**Figure 3**

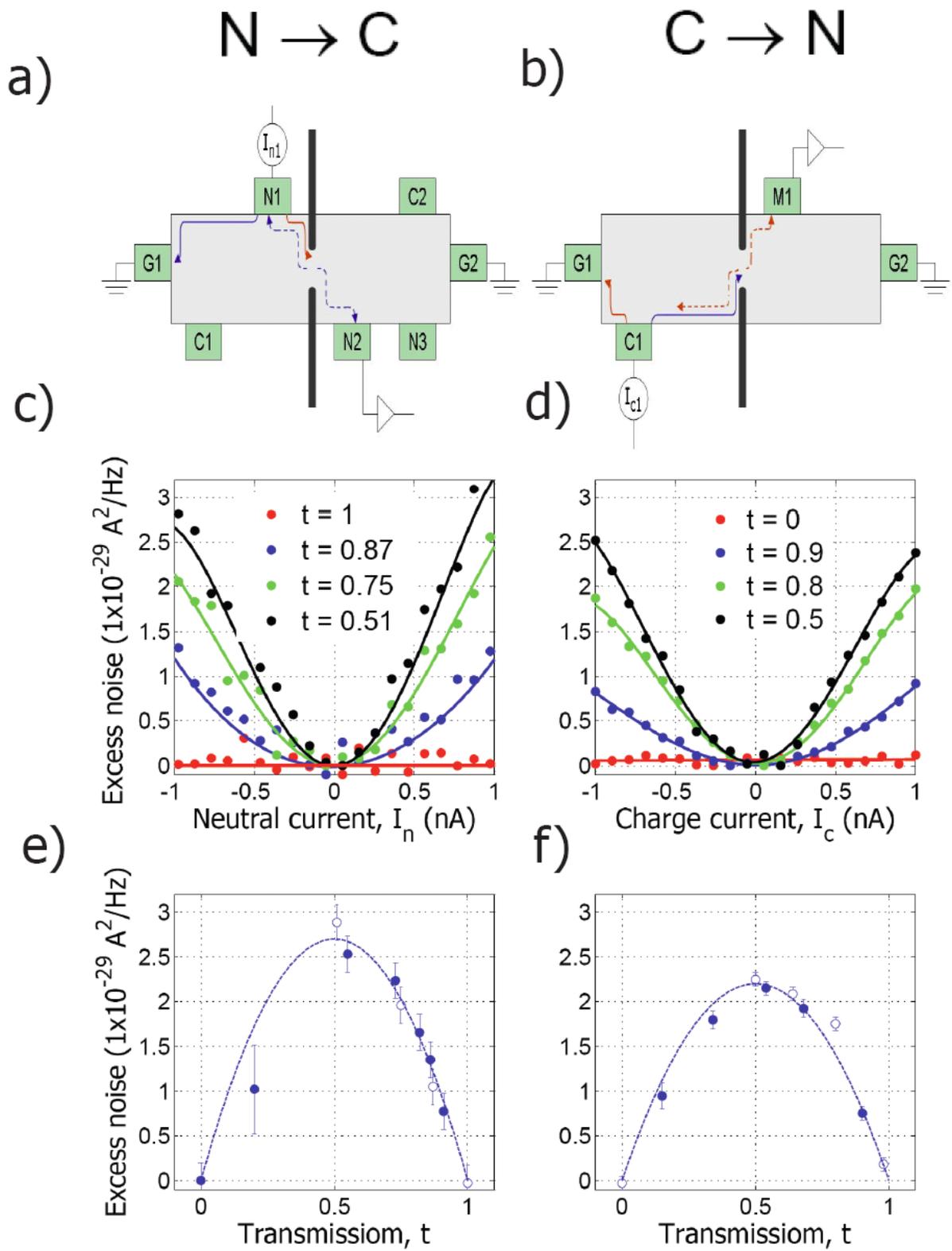

**Figure 4**